\documentclass[%
reprint,
aps,
prl,
amsmath,
twocolumn,
amssymb,
floatfixng,
showpacs,
superscriptaddress,
footinbib,
multicol
]{revtex4-2}
\usepackage{graphicx}
\usepackage{epstopdf}
\usepackage{dcolumn}
\usepackage{bm}
\usepackage{color}
\usepackage{xcolor}
\usepackage{amsmath}
\usepackage{braket}
\usepackage{amsfonts}
\usepackage{soul}
\usepackage{url}
\usepackage{array,tabularx}
\usepackage{graphics}
\usepackage{times}
\usepackage{subfigure}

\usepackage[export]{adjustbox}
\usepackage[colorlinks=true,linktoc=page,linkcolor=red,citecolor=blue]{hyperref}

\definecolor{OliveGreen}{rgb}{0,0.6,0}

\DeclareMathOperator{\sech}{sech}

\begin{document}

\title{
Harnessing Nonlinearity to Tame Wave Dynamics in Nonreciprocal Active Systems
}
\author{Sayan Jana}
\affiliation{School of Mechanical Engineering, Tel Aviv University, Tel Aviv 69978, Israel}
\author{Bertin Many Manda}
\affiliation{School of Mechanical Engineering, Tel Aviv University, Tel Aviv 69978, Israel}
\affiliation{Laboratoire d’Acoustique de l’Universit\'e du Mans (LAUM),
UMR 6613, Institut d’Acoustique - Graduate School (IA-GS), CNRS,
Le Mans Universit\'e, Av. Olivier Messiaen, 72085 Le Mans, France}
\author{Vassos Achilleos}
\affiliation{Laboratoire d’Acoustique de l’Universit\'e du Mans (LAUM),
UMR 6613, Institut d’Acoustique - Graduate School (IA-GS), CNRS,
Le Mans Universit\'e, Av. Olivier Messiaen, 72085 Le Mans, France}
\author{Dimitrios J.  Frantzeskakis}
\affiliation{Department of Physics, National and Kapodistrian University of Athens, Athens 15
784, Greece}
\author{Lea Sirota}\email{leabeilkin@tauex.tau.ac.il}
\affiliation{School of Mechanical Engineering, Tel Aviv University, Tel Aviv 69978, Israel}

\begin{abstract}

We present a mechanism to generate unidirectional  pulse-shaped propagating waves, tamed to exponential growth and dispersion, in active systems with nonreciprocal and nonlinear couplings. In particular, when all bulk modes are exponentially localized at one side of the lattice (skin effect), it is expected that wave dynamics is governed by amplification or decay until reaching the boundaries, even in the presence of dissipation. Our analytical results, and experimental demonstrations in an active electrical transmission line metamaterial, reveal that nonlinearity is a crucial tuning parameter in mediating a delicate interplay between nonreciprocity, dispersion, and dissipation. Consequently, undistorted unidirectional solitonic pulses are supported both for low and high nonreciprocity and pulse amplitude strength.  
The proposed mechanism facilitates robust pulse propagation in signal processing and energy transmission applications.

\end{abstract}

\maketitle


Non-Hermitian systems arising from nonreciprocal interactions have been intensively studied both in theory and experiment~\cite{NatRevMat2020,CalozPRapplied2018}.
Most of the works focused on the linear spectral properties of such systems as a function of the underlying parameters, and in particular on the extension of notions of topology to the non-Hermitian realm~\cite{sato2019,ashida2020non,ninjarev22,fan24}. 
Consequently, unique phenomena was identified, e.g. 
high sensitivity of the spectrum to boundaries, and eigenvectors localization at the system edges, referred to as non-Hermitian skin effect (NHSE)~\cite{OKSS2020,LTLL2023,WC2023}. 
Following these studies, efforts have shifted towards the dynamical features of nonreciprocal lattices, suggesting advanced applications, such as self-healing of modes~\cite{longhi_self}, dynamical NHSE~\cite{DSE22,DSEninja,DSEBEC}, NHSE direction reversal via coherent coupling~\cite{DRSE}, edge burst~\cite{burst22,burstexp}, curved spaces~\cite{lv2022curving}, self-acceleration~\cite{longhi_accel}, energy guiding~\cite{padlewski2024amplitude} and, recently, propagation of nonlinear waves, such as kinks and breathers~\cite{BGVVTCC2024,VGGSMC2024}. 

\begin{figure}[!]
    \centering
\includegraphics[width=\columnwidth]{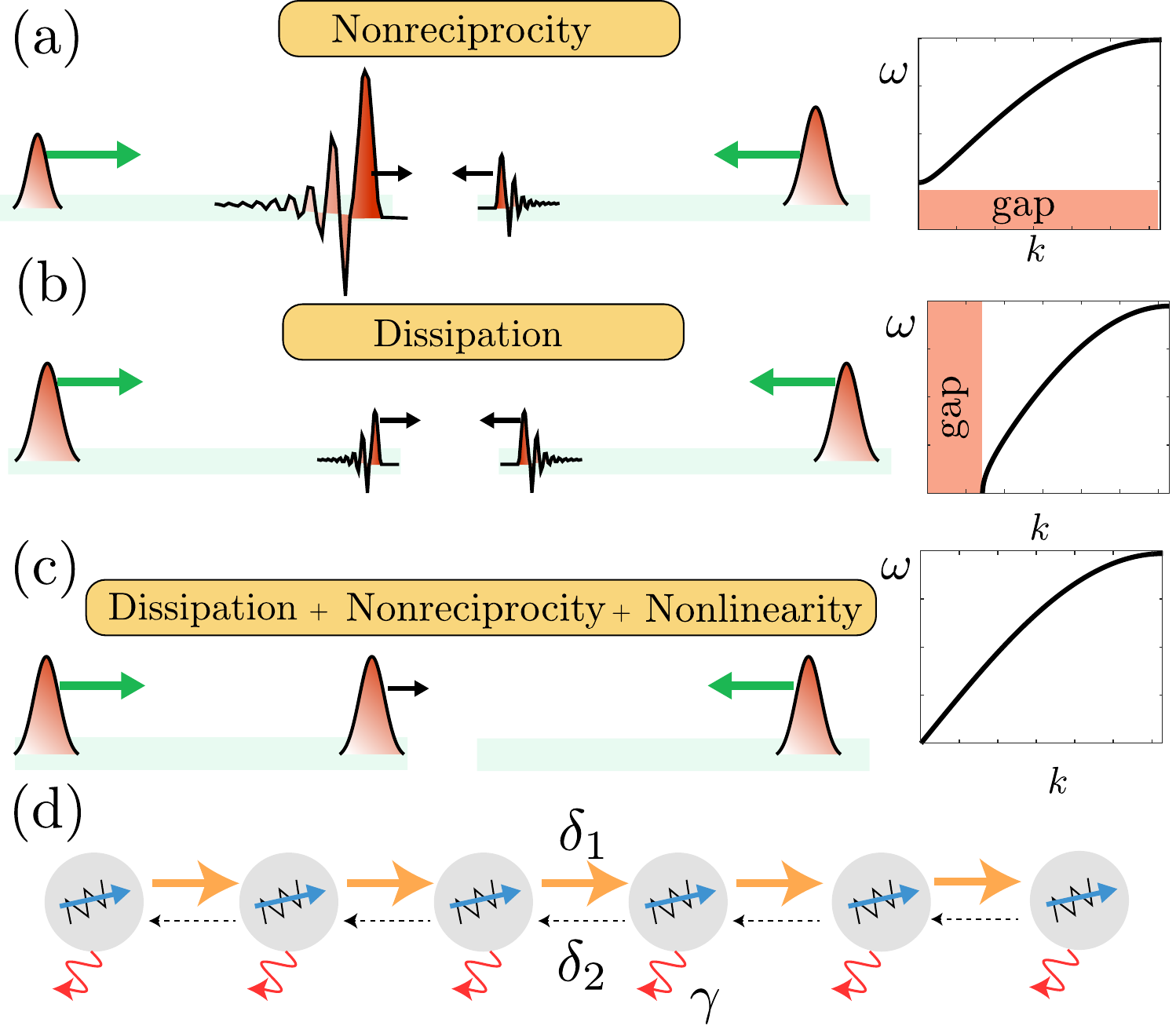}
    \caption{Schematics of pulse propagation in a nonreciprocal (a), a dissipative (b), and a dissipative, nonreciprocal, and nonlinear (c) lattice; right panels in (a)-(c) show the corresponding typical dispersion curves. In case (c), balance of nonlinearity with dispersion, and nonreciprocity with dissipation, leads to the emergence of solitons. Bottom panel (d) depicts schematically the lattice of case (c), with $\delta_{1,2}$ and $\gamma$ denoting couping for right- or left-going waves, and loss. 
     }
\label{fig:gen_scheme}
\end{figure}

A typical characteristic of the NHSE dynamics is the unidirectional wave propagation. Moreover, the dynamics  also feature amplification in one spatial direction
and attenuation in the other, as illustrated in Fig.~\ref{fig:gen_scheme}(a). While the unidirectionality property is desired and exploited in the applications, the amplitude growth is usually a drawback. 
In addition, since the nonreciprocal couplings are usually realized between discrete sites, either in inherent lattices~\cite{rosa2020dynamics}, or in continuous media that is subjected to discrete interaction points~\cite{zhang2021acoustic}, the propagating waves are prone to dispersion. This is also visualized in Fig.~\ref{fig:gen_scheme}(a), where obviously, the narrower is the propagating pulse, the stronger is its dispersion. 

In this work, our goal is to tame the wave dynamics of the NHSE within these two aspects - the spatial amplitude growth and the dispersion of narrow pulses.
For the first aspect, we invoke dissipation, as illustrated in Fig.~\ref{fig:gen_scheme}(b). 
An incoming pulse from either side of a one-dimensional (1D) dissipative lattice is being reduced in amplitude. 
We thus aim at finding a way to tailor the dissipation across the lattice so that to balance the NHSE, and achieve unidirectional waves, in the form of pulses, which do not suffer from exponential growth or decay.

At the next stage we aim at making the unidirectional dynamics dispersionless, as depicted in  Fig.~\ref{fig:gen_scheme}(c), specifically for narrow pulses  compared to the lattice constant. 
For this sake, we invoke nonlinearity, hence considering the pulses as solitons~\cite{solit}.
In reciprocal systems, the interplay between nonlinearity and dispersion is well understood, leading to the development of the theory of nonlinear waves in diverse fields, such as fluids, plasmas, optics, condensed matter physics, and so on~\cite{Dauxois,remoissenet2013waves,Ablowitz,CFK}.
For nonreciprocal lattices, however, this interplay has so far remained an open question.

Here we answer this question both theoretically and experimentally for soliton  that are governed by a generalized version of the Boussinesq equation. The latter was originally proposed as a model for bidirectional shallow water waves~\cite{Ablowitz}, and later extended to plasma physics, optics, mechanical and electrical lattices, and more~\cite{remoissenet2013waves,CFK}. We thus establish connections with different physical contexts, and show that non-Hermitian, nonreciprocal lattices support unidirectional, nonreciprocal solitons that are unique compared to their reciprocal counterparts in various aspects.

As for the experimental realization, nonreciprocal interactions in lattices are usually created using embedded active control elements, which actuate the lattice sites based on real-time measurements of the responses~\cite{geib2021tunable,wen2022unidirectional,wen2023acoustic,jana2023gravitational}. 
In particular in electric circuit lattices, the control process is realized using operational amplifiers \cite{hofmann2019chiral,zanic2022stability,zhu2023higher,jana2023tunneling,benisty2024controlled}, which are active elements that intrinsically stand for both actuators and sensors.
In this work we demonstrate the tamed unidirectional solitons on an electric circuit platform.

{\it Analytical results.} We consider nonreciprocal lattice systems, as schematically depicted in Fig.~\ref{fig:gen_scheme}(d), and described by the dimensionless model
\begin{equation} 
    \begin{split}
        \frac{d^2y_n}{dt^2}=&a_{+}y_{n+1}-2y_n+a_{-}y_{n-1} + F_{\text{dmp}}+ F_{\text{nl}}. 
    \end{split}
    \label{eq:second_order_ODE}
\end{equation}
Here, $y_n(t)$ denotes field amplitude at time $t$ and site $n$ ($n=1, 2, \ldots, N$), whereas the couplings $a_{\pm} = 1 \mp b/2$ ($b>0$) are assumed to be asymmetric, and quantify the nonreciprocity of the system. In addition, $F_{\text{dmp}}$ and $F_{\text{nl}}$ are the local damping and nonlinear forces, respectively. For 
$F_{\text{nl}}=0$, this model is the dissipative analogue of the celebrated Hatano-Nelson model ~\cite{HN1996,HN1998,TRANSFO} with a second derivative in time (as opposed to the Schr\"{o}dinger equation). This analogy is particularly relevant for experimental investigations involving nonreciprocal variations in mass-spring systems~\cite{WWM2022a,WWM2022b,GBVC2020} and LC electric circuits~\cite{A2005}. Hereafter, 
we assume a generic form for the loss, i.e., $ F_{\text{dmp}} = -2 \gamma \frac{dy_n}{dt}$, where $\gamma>0$ is the dissipation strength. 

In the linear limit, small-amplitude solutions are given by:
\begin{align}
y_n(t)= e^{-\gamma t}e^{dn}e^{i\omega t}e^{ikn}, 
\end{align}
where $2d=\ln\left[(1+\frac{b}{2})/(1-\frac{b}{2})\right]$. In particular, due to the $e^{dn}$ term, waves 
grow (decay) for increasing (decreasing) $n$, in a way reminiscent of the NHSE. Furthermore, the solutions also decay in time, due to the $e^{-\gamma t}$ term, regardless of the direction of propagation. 
The corresponding dispersion relation 
\begin{align}
\omega^2 = 2 - 2\sqrt{1 - b^2/4}\cos k - \gamma ^2 \label{linear},
\end{align}
features a gap for a range of low frequencies [Fig.~\ref{fig:gen_scheme}(b)] when dissipation is dominant or, for low wavenumbers [Fig.~\ref{fig:gen_scheme}(a)], when nonreciprocity is stronger. These two regions are separated by the curve $\gamma^2 = 2 - 2\sqrt{1 - b^2/4}$, corresponding to the dispersion relation starting from $\omega=0$ [Fig.~\ref{fig:gen_scheme}(c)]~\cite{supplementary}. 

\begin{figure}[t!]
    \centering
\includegraphics[width=\columnwidth]{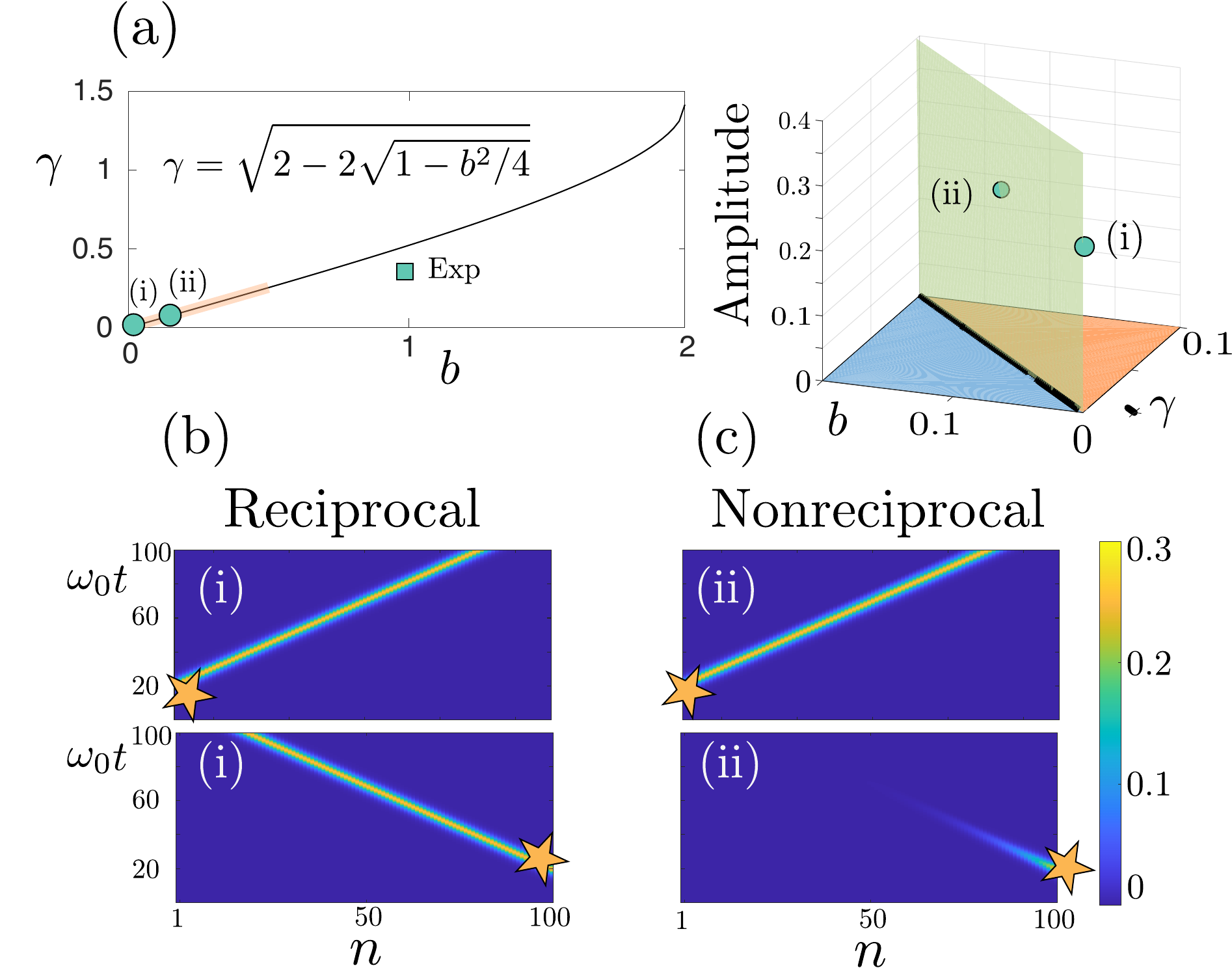}
\caption{
    The nonreciprocity, dissipation and nonlinearity phase diagram, $(b, \gamma, A)$.
    The regions below and above $\gamma^2=2 - 2\sqrt{1-b^2/4}$ (black curve) indicate the nonreciprocal and dissipation dominated phases respectively when $A\rightarrow 0$.
    Right: Family of solitons continuing from weakly dispersive waves of the linearized limit (black curve) when increasing nonlinearity.
    The circle symbols at $b=0$ and $\gamma=0$ (i) and $b=0.1056$ and $\gamma=0.05$ (ii) shows the values for the numerical simulations of soliton propagation, see (c).
    The square symbol represent $b=1$ and $\gamma=0.4$ of the experimental setup.}  \label{fig:theo_num_confirm_soliton}
\end{figure}

Interestingly, for weak nonreciprocity, i.e. $b\ll1$, the spatial growth/decay rate becomes $d=b/2$ and, at the same time, the condition of gap closing approximates to $\gamma=b/2$. In this limit, the latter \emph{balance condition for linear waves} ensures ---at the same time--- that the gap vanishes and the decay is balanced by the nonreciprocity. 
In Fig~\ref{fig:theo_num_confirm_soliton}(a), we plot the curve $\gamma=\gamma(b)$ on which the gap vanishes with the region of the linear balance indicated by the thick line near the origin. 
Alternatively, traveling wave solutions, in the form of $y=y(s)$ with $s=n \pm t$  may be sought. In this case, both such right- and left-going waves decay for $b<2\gamma$.  
On the other hand, right- (left-) going waves are expected to grow (decay) when traveling to the right (left)  for $b>2\gamma$.
At the critical value 
$b=2\gamma$ loss is compensated by nonreciprocity suggesting loseless pulse propagation.
It is in this parameter regime that we seek for
soliton solutions.

Next, we proceed with the nonlinear regime, and investigate the emergence of traveling nonlinear waves. To a first approximation, a weak (quadradic) nonlinearity, pertinent to our experimental setup (see below), is taken to be $F_{\text{nl}} = \beta \frac{d^2}{dt^2}\left(y_n^2 \right)$ (a similar form also appears in other systems, such as acoustic waveguides~\cite{IOANNOU}). Then, 
in the long wavelength limit ($k\ll 1$), and for weak nonreciprocity and dissipation ($b,\gamma\ll 1$), the continuum approximation, $y_n(t) \rightarrow y(x=n, t)$, for  Eq.~(\ref{eq:second_order_ODE}) is the following nonlinear partial differential equation:
\begin{equation}
\!\!\!
    \frac{\partial^2 y}{\partial t^2} - \frac{\partial^2 y}{\partial x^2} - \frac{1}{12}\frac{\partial^4 y}{\partial x^4} - \beta\frac{\partial^2}{\partial t^2}\left( y^2\right) = -b \frac{\partial y}{\partial x} - 2\gamma \frac{\partial y}{\partial t}.
\label{eq:second_order_ODE_continuous}
\end{equation}
%
Equation~\eqref{eq:second_order_ODE_continuous} is interesting on its own right: for $b=\gamma=0$ it reduces to the standard Boussinesq equation, a classical bidirectional shallow water wave model~\cite{Ablowitz}, while in our case ($b,\gamma \neq 0$) it incorporates ---apart from the linear loss--- a nonreciprocity term. Note that a similar equation was derived in Refs.~\cite{VGGSMC2024,BGVVTCC2024} but with an additional ``mass'' term, $\propto y$, which inevitably forms a gap and forbids the propagation of pulses.

\begin{figure*}[htb!]
    \centering
\includegraphics[width=17cm]{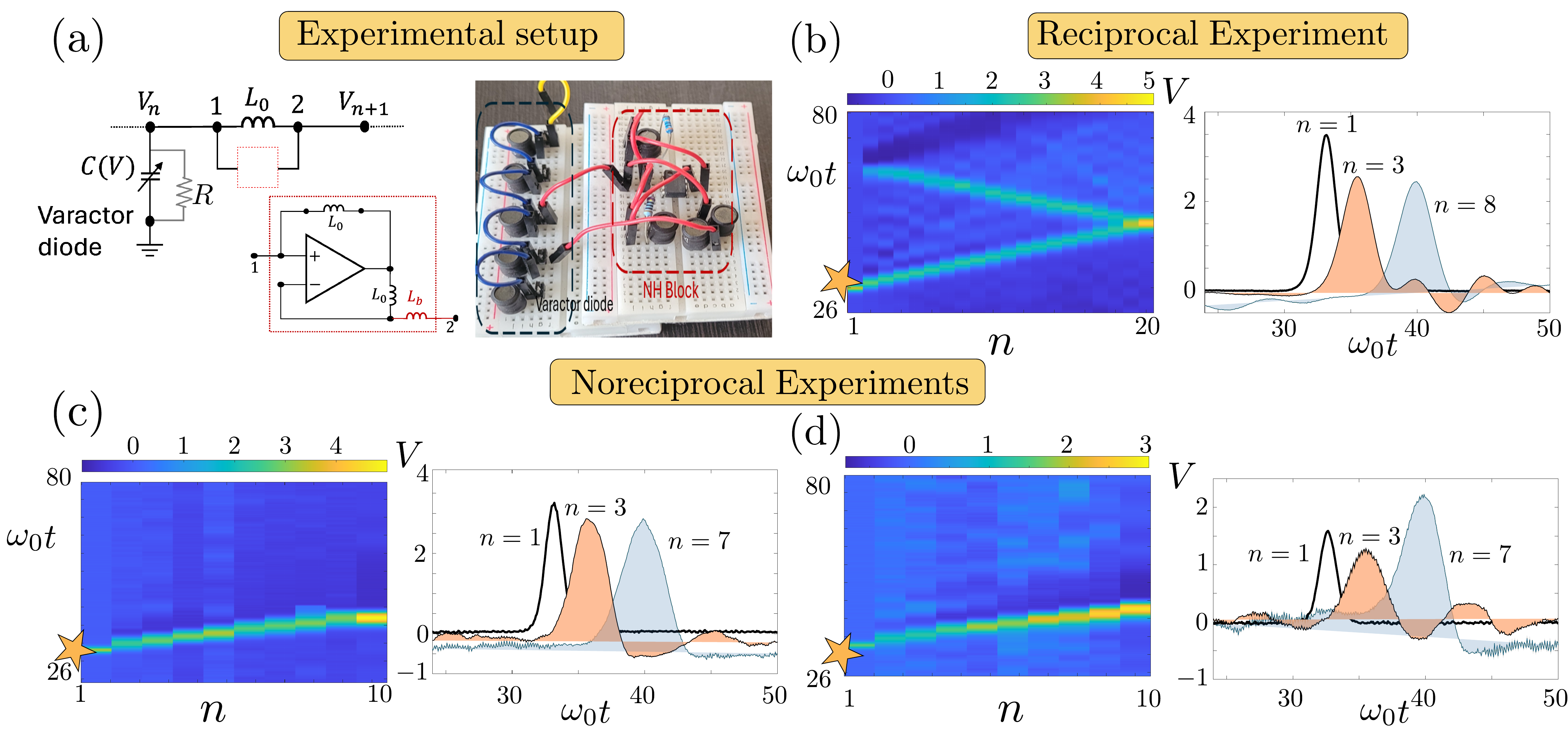}
    \caption{
    (a) Active circuit implementation of  nonlinear nonreciprocal and dissipative electric metamaterials.
    The nonlinearity is induced through a voltage dependent capacitor (varactor diode).
    The nonreciprocity is generated through a voltage supplied OpAmp, which breaks the reciprocity of current flow through the unit cell. (b)
    Spatio-temporal voltage response to an input pulse, $V_s(t)$, of amplitude $A$ and width $\tau$ at the left of the chain (star symbols).
    In panel (b) (reciprocal experiment,  $b=0$), while in panels (c) and (d) (nonreciprocal experiments, $b=1$) we use $A=4~\mbox{V}$ and $A=2~\mbox{V}$, respectively; in all cases $\tau=0.66~\mbox{$\mu$s}$. 
    }
\label{fig:experiment_all_included}
 \end{figure*}
Following the intuition from the linear analysis, we now seek propagating solutions in the from of nonlinear pulses for low wavenumbers. We thus look for solutions of Eq.~\eqref{eq:second_order_ODE_continuous} of the form $y=y(s)$, where $s=x \pm vt$ ($v$ is the nonlinear wave velocity), 
having the profile of Boussinesq solitons~\cite{supplementary}, 
\begin{align}  \label{eq:the_soliton}
y(x,t)=\frac{3\left(v^2-1\right)}{2\beta v^2}\sech^2\left[\sqrt{3\left(v^2-1\right)}\left(x-vt\right)\right], 
\end{align}
whenever 
\begin{align}  
     v= b/2\gamma.
     \label{nlbc}
\end{align}
Equation~(\ref{nlbc})  constitutes a \emph{balance condition for nonlinear waves}, and predicts \emph{uni-directional} soliton propagation,  
dictated by the sign of the nonreciprocity parameter $b$. 
Recalling that for a fixed value of $b/2\gamma>1$ linear pulses are expected to (grow) decay and also disperse, we can now appreciate the role of  
nonlinearity, which can 
balance all these effects, and allows for the propagation of solitons. The latter are 
of fixed amplitude $A= 3(b^2-4\gamma^2)/2\beta b^2$, as dictated by Eq.~(\ref{nlbc}), which 
``locks'' the soliton parameters (for given values of $b$ and $\gamma$), contrary to what is commonly known in Hermitian systems, where the amplitude (or the velocity) is a free parameter.
Hence, the obtained soliton is reminiscent to ''auto-solitons'' occurring in other dissipative systems as, e.g., the complex Ginzburg-Landau equation \cite{GL}.

Nevertheless, this soliton is not isolated, as it belongs to a family of solutions lying on the plane  $A=3(b^2-4\gamma^2)/2\beta b^2$ of the $(b, \gamma, A)$ extended parameter space 
--see right panel of Fig.~\ref{fig:theo_num_confirm_soliton}(a). In other words, our solutions extend the usual bidirectional Boussinesq solitons lying on the line $b=\gamma=0$, to a family of unidirectional ones lying on the respective surface.

Our analytical predictions are confirmed by direct numerical simulations~\cite{DNS2024} performed at the level of 
the discrete problem, for a lattice of $N=100$. We present results for solitons of amplitude $A=0.254$, using time-dependent boundary condition of the form of Eq.~\eqref{eq:the_soliton} at either the left ($n=0$) or the right $(n=N+1)$ boundary. 
For illustration reasons, first we show a bidirectional soliton in the dissipationless and reciprocal case ($b=\gamma=0$) in Fig.~\ref{fig:theo_num_confirm_soliton}(b) (star symbols indicate the source). This simulation recovers the behavior of the well-known 
pulse-shaped solitons of the Boussinesq equation~\cite{CFK} 
which travel undistorted along the lattice (here we choose $\beta=0.23$). 
On the other hand, when both nonreciprocity and dissipation are present, unidirectional solitons aligned with the amplifying direction persist, and travel undistorted for a long distance, in accordance with our predictions 
---see Fig.~\ref{fig:theo_num_confirm_soliton}(c). Contrary, as shown in the same figure, if the source is located at the opposite side, both nonreciprocity and dissipation act in unison, and the pulse 
undergoes a fast decay. 

\paragraph*{Experimental results.}
To demonstrate these results experimentally, we construct an active electric transmission line in which each unit cell is realized by nonlinear 
capacitors with $C(V)=C_0\left[(1-C_e)(1+V/V_j)^{-m}+C_e\right]$,  
where $C_0=102~\mbox{pF}$, $C_e=0.1$, $V_j=25~\mbox{V}$ and $m=12.76$, connected via inductors $L_0 = 220~\mu\text{H}$,
as indicated in Fig.~\ref{fig:experiment_all_included}(a). 
Active NH blocks, formed by OpAmps in a feedback configuration, are attached in parallel to the inductors, and break the reciprocity of the current flow between the unit cells as $L_0\dot{I}_{n\rightarrow n+1}=(1-L_0/L_b)\Delta V$ and $L_0\dot{I}_{n+1\rightarrow n}=(1-L_0/L_b)/\Delta V$, with $L_b=L_0/(b/2)$, and $\Delta V$ being the voltage drop across the unit cell~\cite{jana2023tunneling}.
The voltage response $V_n(t)$ is then governed by
\begin{align}
    L_0\frac{d}{dt}\left[C(V_n)\frac{d V_n}{dt}\right]=a_{+}V_{n+1}-2V_n+a_{-}V_{n-1} -\frac{1}{R}\frac{ d V_n}{dt}.    \label{eq:full_circuit_equations}
\end{align}
For sufficiently small voltages, we use 
$C(V) \approx C_0(1-2\beta V)$, 
and by setting $t \rightarrow \omega_0 t$ ($\omega_0 = 1/\sqrt{L_0C_0}$) and $R\rightarrow (1/2\gamma)\sqrt{L_0/C_0}$, we recover the model of Eq.~\eqref{eq:second_order_ODE}.
We use input pulse signals of the form
$V_s(t) = A\sech^2 \left[\left(t - t_0\right)/\tau  \right]$ at the left of the chain, with $A$, $\tau =0.66~\mbox{$\mu$s}$ and $t_0$ being respectively the amplitude, 
width and center 
of the input signal.
A first  experiment is conducted in the absence of the OpAmps, i.e., in 
the dissipationless reciprocal chain of $N=20$ cells.
This setup corresponds to  $b=\gamma=0$ 
indicated by a dot symbol in Fig.~\ref{fig:theo_num_confirm_soliton}(a). As shown in Fig.~\ref{fig:experiment_all_included}(b), the initial pulse amplitude is equal to $A=4~\mbox{V}$~\cite{noise}, beyond the small-amplitude approximation. However, it is well known~\cite{Ablowitz,CFK}, that the wave initially radiates and eventually reorganizes itself into a soliton, propagating almost undistorted, as shown by the experimentally obtained spatiotemporal evolution, and measured time signal at different cells.

The next set of experiments is done using the OpAmps in a smaller lattice of $N=10$, and with our chosen inductances,  $L_b=2L_0=440~\mbox{$\mu$H}$,   we achieve strong nonreciprocity with $b=1$. 
In addition, the whole setup is inherently dissipative, with losses primarily arising from the experimental components, high-frequency 
OpAmp–inductor–varactor connections, and various point contact connections throughout the system. To estimate the losses we have performed simulations using the SPICE software~\cite{NP1973, N1975, webSPICE} adding resistance $R$ at each unit cell and adjusting its value so that  experiment and simulation agree~\cite{supplementary}; this way, we have obtained a value of  $\gamma\approx 0.4$ [see 
square symbol in Fig.~\ref{fig:theo_num_confirm_soliton}(a)].

As such, it follows that the system lies within the nonreciprocity dominated phase and
traveling pulses will generically grow. Although the parameters are not in the region of validity of Eq.~\eqref{eq:second_order_ODE_continuous}, we find [see Fig.~\ref{fig:experiment_all_included}(c)]
that for the same input amplitude as above, $A=4~\mbox{V}$, the nonlinearity prevents the pulse from dispersing~\cite{supplementary}, and a soliton traveling with practically constant amplitude up to $\omega_0 t=50$ is formed.
Further, a striking difference compared to the experiment of the reciprocal lattice emerges: once the pulse reflects at the system's right boundary [see bright spot at $n=10$ in Fig.~\ref{fig:experiment_all_included}(c)] it decays exponentially fast.

For the same experiment, but with 
an input pulse of amplitude $A=2~\mathrm{V}$, the corresponding result
is shown in Fig.~\ref{fig:experiment_all_included}(d).
We now observe wave amplification, with the output signal reaching amplitude of $\approx 3~\mbox{V}$.
To further clarify the dynamics, we plot the voltage profiles at 
$n=1$, $3$ and $7$, as before. 
The input pulse at 
$n=1$ [black curve of Fig.~\ref{fig:experiment_all_included}(d)] radiates at the early stage of its evolution, and as a result its amplitude decays --see orange filled pulse at $n=3$ in Fig.~\ref{fig:experiment_all_included}(d). Contrary to the $A=4~\mathrm{V}$ case, the obtained pulse fails to maintain its shape, and its amplitude at $n=7$ becomes twice as large 
--see blue filled pulse in Fig.~\ref{fig:experiment_all_included}(d).

\begin{figure}[t!]
    \centering
    \includegraphics[width=8.4cm]{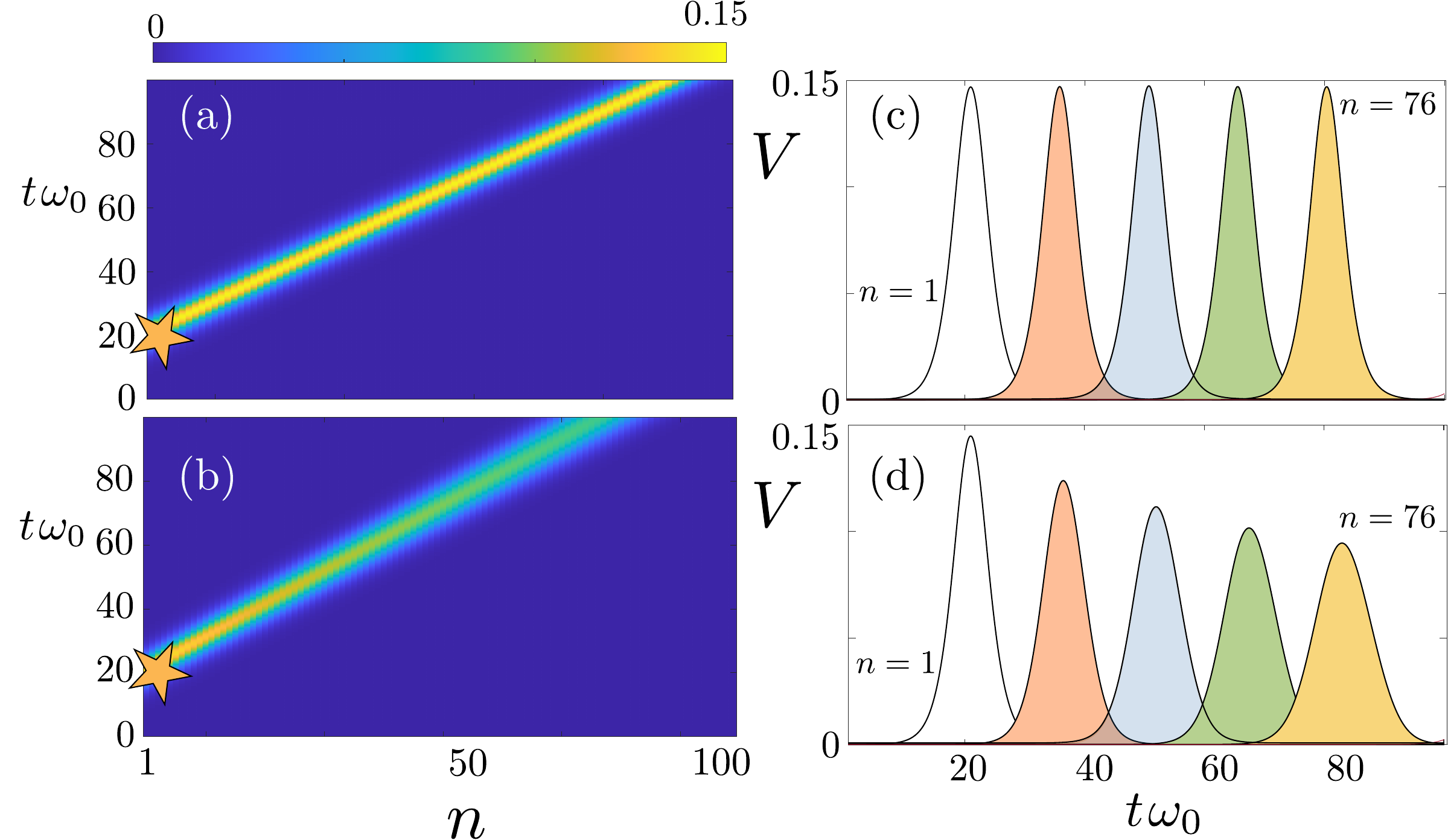}
    \caption{
        Spatiotemporal voltage response to a soliton-like pulse excitation  of amplitude $A=0.15$ (star symbols) at the left of the chain  of length $N=100$ in the nonreciprocal dominated phase with (a) $b=1$, $\gamma_0 =0.5176$ $\gamma_1 =0.232$ ($\gamma=0.48$).
        (b) Same as (a) but for $\gamma_1 = 0$.
        (b) Snapshots of (a) depicting the time dependence of the voltage response at site $n=1$, $16$, $31$, $46$, $61$ and $76$ from the left to right respectively.
        (d) Same as (c), but for $\gamma=0$.
    }
    \label{fig:simulations_nonlinearloss}
\end{figure}

We further compare 
the experimental results with our analytical predictions,
which are disconnected in the parameter space of Fig.~\ref{fig:theo_num_confirm_soliton}(a). 
Indeed, the theory predicts the existence of weak solitons in the regime of small nonreciprocal strength.
On the other hand, we conducted experiments for large-amplitude, nonreciprocal strength, and pulse excitations.
We thus point out that it is possible to find traveling solitons, in both theory and experiment, for large values of the nonreciprocity strength~\cite{REQUIREMENT}.

To do so, we apply a tweak to our original apparatus, introducing a voltage-dependent resistor (known as ``varistor''~\cite{S1951}) at each lattice site, following $F_{\text{dmp}}(y)=-2\frac{d}{dt}\left(\gamma y\right)$ where the resistance expands as $\gamma = \gamma_0 - \gamma_1 y + \mathcal{O}\left(y^2\right)$.
We find that, in the continuum limit, the system can be described, to the leading-order of approximation, by the
Boussinesq equation~\eqref{eq:second_order_ODE_continuous}, but with the convection-diffusion-like term leaving its place to a Korteweg-De Vries-like counterpart, through the addition of $-\frac{b}{6}\frac{\partial^3 y}{\partial x^3} + 2\gamma_1 \frac{\partial y^2}{\partial t}$~\cite{supplementary}. In this context, we find robust traveling nonreciprocal soliton states, 
provided that $2bv = \gamma_0 + \sqrt{2b^2 + \gamma_0^2}$ and $\gamma_1 = b\beta v$ (with $b>2\gamma_0$), which is a \emph{modified} balance condition for nonlinear waves.

Figure~\ref{fig:simulations_nonlinearloss} depicts representative cases of the dynamical response to an input pulse [Eq.~\eqref{eq:the_soliton}] of amplitude $A=0.15$, when $b=1$.
Clearly, in the presence of nonlinear resistance, and specifically for  $\gamma_1=0.232$ (i.e., $\gamma\approx 0.48$), the input pulse travels at constant amplitude, width and velocity ---see Figs.~\ref{fig:simulations_nonlinearloss}(a),(c)-- featuring the behavior of a genuine soliton.
On the other hand, for $\gamma_1=0$, the initial pulse widens and decays during its evolution --see Figs.~\ref{fig:simulations_nonlinearloss}(b),(d).
Nevertheless, no dispersion is observed, as radiating tails are absent. 

In conclusion, we demonstrated the formation and dynamics of a family of unidirectional Boussinesq type solitons in nonlinear lattices with nonreciprocal couplings. 
We showed, both theoretically (for weak nonlinearity), and experimentally in active transmission lines (for strong nonlinearity), that the amplitude of an input pulse enables the balance between dissipation, dispersion, and nonreciprocity, thus taming the NHSE wave dynamics. 
In addition, we analytically showed that robust propagation of narrow nonlinear pulses is possible for strong nonreciprocity with the inclusion of a nonlinear loss term and a resulting modified balance condition. 
We anticipate that this work will serve as a foundation for exploring robust transport in nonlinear dissipative and active systems.

\section*{Acknowledgements}

\textit{Funded by the European Union. Views and opinions expressed are however those of the author(s) only and do not necessarily reflect those of the European Union or the European Research Council Executive Agency. Neither the European Union nor the granting authority can be held responsible for them. V.A and B.M.M are supported by ERC grant NASA -  101077954. S.J. and L.S. are supported in part by the Israel Science Foundation Grants No. 2177/23 and 2876/23. 
The authors are grateful to Arkadi Rafalovich, Gil Koresh, Johnny Benisty, and Sanskruti Smaranika Dani for invaluable assistance with the experimental setup.}

\bibliography{nonlinear}

\begin{thebibliography}{56}
\bibitem [{sol()}]{solit}%
  \BibitemOpen
  \href@noop {} {}\bibinfo {note} {Hereafter, we use the term soliton in a loose sense without implying complete integrability \cite{Ablowitz}.}\BibitemShut {Stop}%
\bibitem [{TRA()}]{TRANSFO}%
  \BibitemOpen
  \href@noop {} {}\bibinfo {note} {The transformation $\phi_n=e^{i\sqrt{2}t}y_n$ can further be applied to remove the onsite potential.}\BibitemShut {Stop}%
\bibitem [{noi()}]{noise}%
  \BibitemOpen
  \href@noop {} {}\bibinfo {note} {Due to the high sensitivity of our experimental platform to noise in a low voltage regime.}\BibitemShut {Stop}%
\bibitem [{REQ()}]{REQUIREMENT}%
  \BibitemOpen
  \href@noop {} {}\bibinfo {note} {In our current experimental setup, advanced numerical continuation methods are the only feasible approach to extend the Boussinesq family of solitons from small non-reciprocal strengths to higher ones.}\BibitemShut {Stop}%
\end{thebibliography}

\onecolumngrid

\appendix

\renewcommand{\thefigure}{S\arabic{figure}}
\renewcommand{\theequation}{S\arabic{equation}}
\setcounter{equation}{0}

\newpage

\section*{\textbf{Supplementary Material}}

\section{Additional experimental characterizations}

\subsection{Varactor diode characterization}

\begin{figure}[htb!]
    \centering
\includegraphics[width=0.3\textwidth]{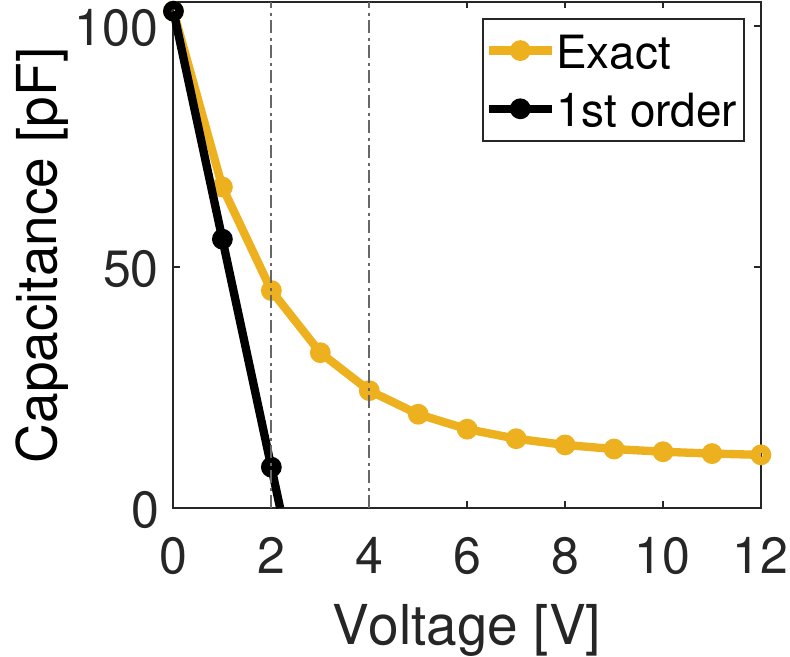}  
    \caption{
Nonlinear capacitance as a function of voltage, as implemented by 
silicon junction varactor diodes with base capacitance $C_0=102~\mathrm{pF}$. Shown are both the fully nonlinear dependence  (yellow curve) and its first order approximation (black line).
    }
   \label{fig:nonlinear_capacitance}
\end{figure}

In our electric circuit unit cell, we utilize a voltage-dependent nonlinear capacitance implemented by a silicon varactor diode. According to the vendor's technical description, the general form of the capacitance is expressed as:
\begin{equation}
    C(V)=C_0\left(\frac{1-C_e}{(1+V/V_j)^m}+C_e\right),
\end{equation}
 where $C_0=102~\mbox{pF}$, $C_e=0.1$, $V_j=25~\mbox{V}$ and $m=12.76$.
The dependence of the cappacitance on the voltage is shown in 
Fig.~\ref{fig:nonlinear_capacitance}  with the yellow curve. The first-order approximation, valid for a small-amplitude voltage (this assumption is used in our analysis), is given by:
\begin{equation}
       C(V) = C_0\left(1 - 2\beta V\right), \quad \beta = \frac{m(1-C_e)}{2V_j},
\end{equation}
and it is also depicted in Fig.~\ref{fig:nonlinear_capacitance} with the black line.

\begin{figure}[ht!]
    \centering
    \begin{tabular}{c}
         (a)  \\
        \includegraphics[width=0.5\columnwidth]{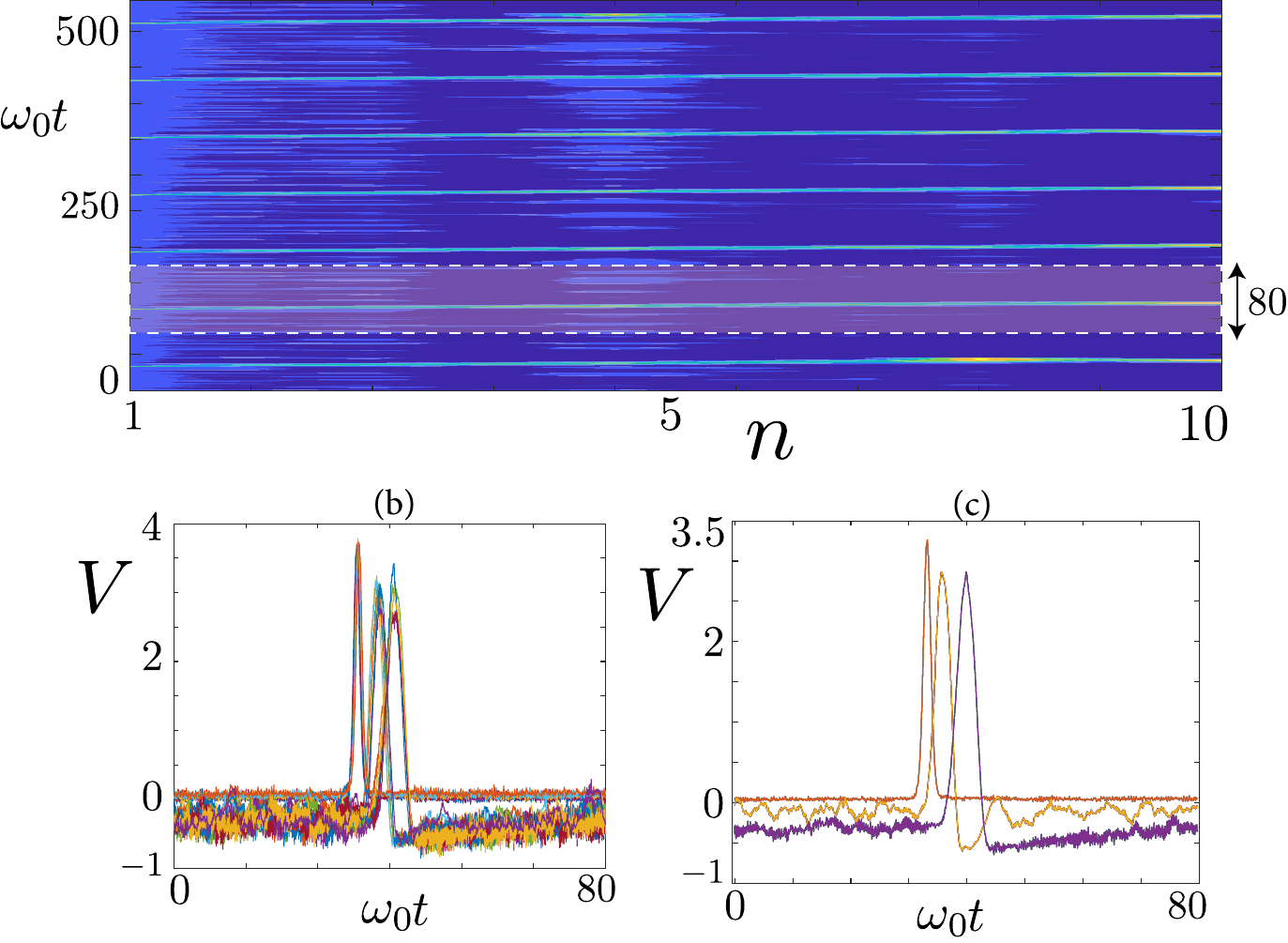}
    \end{tabular}
    \caption{
Experimental data acquisition and analysis, illustrated by the propagation of an input pulse with amplitude $A=4~\mbox{V}$ on a 
lattice of $N=10$ cells,  with $b=1$ and $\gamma \approx 0.4$. 
(a) Response of the system to a train of $7$ input pulses (see text for details).
(b) The oscillogram is divided into $6$ windows of $\omega_0 t = 80$, each capturing the dynamics of a single pulse, which are then overlapped. The time evolution of the pulse at $n=1$, $n=3$, and $n=7$ is depicted from left to right, respectively.
(c) Mean values of the pulse dynamics presented in (b).
    }
\label{fig:experimental_analysis}
\end{figure}

\subsection{Experimental data processing}

To excite the system we use a time-dependent voltage signal composed of $6$ identical consecutive pulses at the beginning of the lattice. These pulses are separated by a time delay of $\omega_0t\approx 80$. 
The system response is recorded at each lattice cell, and an example of the total spatio-temporal result is shown in the top panel of Fig.~\ref{fig:experimental_analysis}. T
To reduce the noise,  
we extract the dynamics of individual input pulses by decomposing the oscillogram into identical windows of width $\omega_0 t = 80$;  these $5$ pulses are shown superimposed on each other in the bottom left panel of Fig.~\ref{fig:experimental_analysis}. Then, taking the average over the $5$ pulses, results in the final signal shown in the bottom right panel.

\subsection{Experimental characterization of the dissipation}

\begin{figure}[hb!]
\begin{tabular}{c c}
(a)  &   (b) \\
\includegraphics[height=4cm, width=5cm, valign=c]{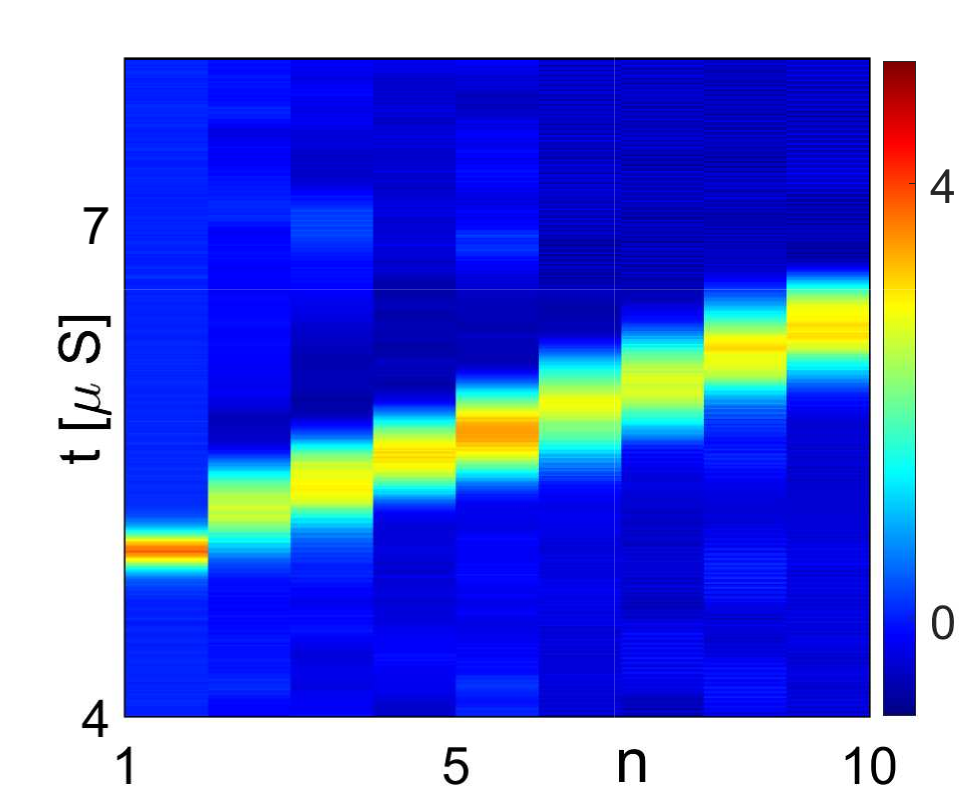}  &   \includegraphics[height=4cm, width=5cm, valign=c]{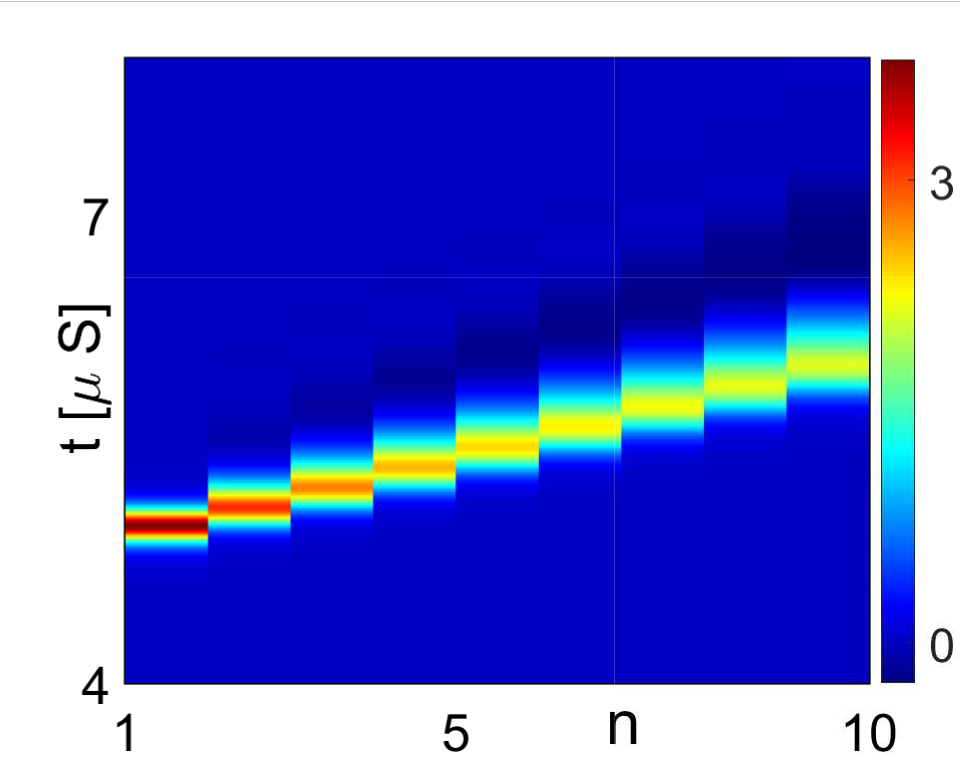} 
\end{tabular}
\caption{Spatio-temporal voltage response to an input pulse, \(V_s(t)\), of amplitude $A=4~\mbox{V}$ and width $\tau=0.66~\mu\mathrm{s}$, applied to the left end of the lattice.
Panel (a) shows the nonreciprocal experimental result, while panel (b) presents the nonreciprocal SPICE simulation result with an onsite resistor value of $R=1850$~$\Omega$. In both cases, we use $b=1$. 
}
\label{fig:spice}
\end{figure}

The system is naturally inherently dissipative, with losses primarily arising from the experimental components, high-frequency op-amp–inductor–varactor connections, and various point contact connections throughout the system. 
Our modeling however incorporates an onsite resistor ($R$) at each node. In order to justify our approximation and also quantify the value of the resistane $R$ we employed a SPICE simulator which models the experimental setup using the same parameters and input data as in the actual experiment. Then we have found the value of $R$ that replicates as close as possible the experimental results.  A resistance value of \(1850~\Omega\) was found to provide a good match with the experimental dynamics as demonstrated in the spatio-temporal response, with the results from the SPICE simulation and experimental data shown in Fig.~\ref{fig:spice}(a) and (b), respectively.

\subsection{Demonstration of taming the dispersion by nonlinearity}

\begin{figure}[htpb]
    \centering
\includegraphics[height=4.5cm]{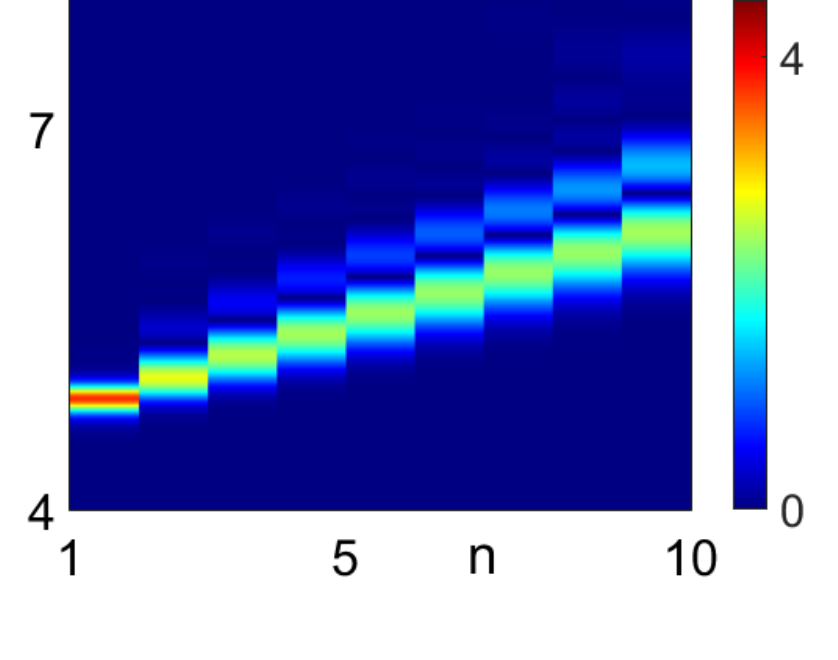}  
    \caption{Numerical simulation of the linear nonreciprocal chain ($\gamma=0.4$, $b=1$), when the varactors are replaced by regular ceramic capacitors of $C_0=102~\mathrm{pF}$. Dispersion of the pulse is observed.
    }
   \label{fig:dispersion_taming}
\end{figure}

To further elaborate on the results in Fig. 3(c),(d) of the main text, it is demonstrated in Fig. \ref{fig:dispersion_taming} that when the nonlinearity is switched off, the pulse begins to disperse. This confirms the necessity of the nonlinear component in taming the NHSE with respect to dispersion, and contributing to the overall balance of nonreciprocity, dissipation, and nonlinearity.

In addition, for $A=4~\mathrm{V}$ the pulse in Fig. 3(c) of the main text traverses the lattice at $\approx 1.1~\mathrm{\mu s}$, compared to $\approx 1.5~\mathrm{\mu s}$ for $A=2~\mathrm{V}$, Fig. 3(d) of the main text, confirming the nonlinear nature of the system through the dependence of the velocity $v$ on the amplitude $A$.

\section{\label{sec:theo}Theoretical analysis}

We recall that the differential-difference equation describing the bulk dynamics of the electric metamaterials is of the form: 
\begin{equation}
    \frac{d^2y_n}{dt^2}  = a_{+} y_{n+1} - 2 y_n + a_{-}y_{n-1} - 2\gamma \frac{dy_n}{dt} + \beta \frac{d^2}{dt^2}\left( y_n^2\right),
    \label{eq:ODEs}
\end{equation}
where $\beta$ is the nonlinear parameter, while $ a_\pm = 1 \mp b/2$, 
and we assume, without loss of generality, that $b>0$ and $\gamma>0$.

\paragraph{Dispersion relation.}
To derive the dispersion relation, we consider the linear limit ($\beta \rightarrow 0$), 
and assume solutions of the form,
\begin{equation}
    y_n(t) = y_0 e^{-\gamma t}e^{i\omega t} e^{dn}e^{-ikn}, \quad d = \ln \left(\sqrt{\frac{a_{-}}{a_{+}}}\right),
    \label{eq:ansatz}
\end{equation}
where $y_0={\rm const.}$ Substituting, we obtain  the following dispersion relation,
\begin{equation}
    \omega^2 = 2 - 2\sqrt{1 - \frac{b^2}{4}}\cos k - \gamma ^2 .
    \label{eq:dispersion_k_to_w}
\end{equation}
Note that, for long waves, $k \ll 1$, we use the approximation 
$\cos k \approx 1$, and obtain:
\begin{align}
     \omega^2 \approx 2 - 2\sqrt{1-\frac{b^2}{4}}-\gamma^2+\left( 2\sqrt{1-\frac{b^2}{4}} \right)k^2.
\end{align}
Hence, the condition $\gamma_c ^2 = 2 - 2\sqrt{1-\frac{b^2}{4}}$ ensures the closing of the gap at low frequencies.

\paragraph{Continuous model.}

To derive the continuous approximation we use   $n\rightarrow x$, $y_n (t) \rightarrow y(x, t)$, and the Taylor expansion:
\begin{equation}
    \begin{split}
            y_{n\pm1}(t) &= y(x, t) \pm \frac{1}{1!}\frac{\partial y(x, t)}{\partial x} + \frac{1}{2!}\frac{\partial^2 y(x, t)}{\partial x^2}   
             \pm \frac{1}{3!}\frac{\partial^3 y(x, t)}{\partial x^3} + \frac{1}{4!}\frac{\partial^4 y(x, t)}{\partial x^4} + \ldots  \\ 
    \end{split}
\end{equation}
We then substitute the expressions above into the discrete equations of motion [Eq.~\eqref{eq:ODEs}] and obtain a continuum model, namely the generalized Boussinesq equation: 
\begin{equation}
    \frac{\partial^2 y}{\partial t^2} - \frac{\partial^2 y}{\partial y^2} - \frac{1}{12}\frac{\partial^4 y}{\partial x^4} - \beta \frac{\partial^2y^2}{\partial t^2} = -b \frac{\partial y}{\partial x} - 2\gamma \frac{\partial y}{\partial t},
    \label{eq:continuum_small_b_01}
\end{equation}
where we have assumed that $b$ is small and, hence, higher-order odd spatial derivative terms can be neglected.
\paragraph{Soliton solutions}
We look for traveling wave solution $y(s)$,  where $s=x\pm vt$. 
Upon substituting into the continuum model, we obtain:
\begin{equation}
    v^2 \frac{d^2y}{ds^2} - \frac{d^2y}{ds^2} - \frac{1}{12}\frac{d^4y}{ds^4} - \beta v^2 \frac{d^2y^2}{ds^2} + \left(b \pm 2\gamma v \right)\frac{dy}{ds} =0.
   \label{eq:moving_frame}
\end{equation}
The above equation defines  an effective dissipative dynamical system owing to the first-order derivative in the equation of motion.
Interestingly, for left- and right-going waves we respectively have:
\begin{equation}
    \begin{cases}
        s = x + vt & \Rightarrow  \left(b + 2\gamma v\right)\frac{dy}{ds}, \\ 
        s = x - vt & \Rightarrow  \left(b - 2\gamma v\right)\frac{dy}{ds}. \\ 
    \end{cases}
\end{equation}
Hence, for $b>0$ and $\gamma >0$, the dissipative term $\propto dy/ds$ in Eq.~\eqref{eq:moving_frame} vanishes only for right-going waves 
provided that:
\begin{equation}
    b = 2\gamma v.
\end{equation}
The latter expression is different from the corresponding for linear waves ($b=2\gamma$). As such, it shows that amplitude itself can be used here as a tuning parameter to achieve seemingly conservative dynamics in one direction. On the other hand, it also locks the velocity of the soliton according to the values of the system parameters $b$ and $\gamma$.
The resulting effective model is conservative, defined by
\begin{equation}
    v^2 \frac{d^2y}{ds^2} - \frac{d^2y}{ds^2} - \frac{1}{12}\frac{d^4y}{ds^4} - \beta v^2 \frac{d^2y^2}{ds^2} = 0,
\end{equation}
and admits a Boussinesq-type soliton of the form~\cite{R1993}:
\begin{equation}  \label{eq:the_soliton_1}
y_n(t)=\frac{3\left(v^2-1\right)}{2\beta v^2}\sech^2\left[\sqrt{3\left(v^2-1\right)}\left(n-vt\right)\right].
\end{equation}

\section{Numerical scheme}

\begin{figure}[h!]
    \centering
\includegraphics[width=0.35\columnwidth]{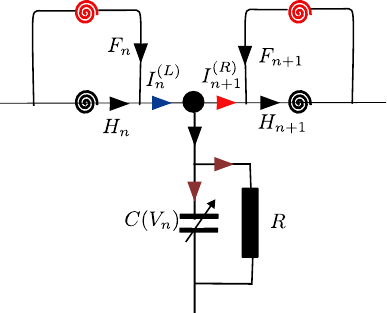}
    \caption{The symbolic unit cell current circulation of the numerical scheme of the electric circuit in the main text.}
    \label{fig:numerical_scheme_unit_cell}
\end{figure}

\bigskip

The lattice simulations and the results shown in Fig.~2 are obtained as follows. We consider Kirchhoff's laws:
\begin{align}
V_{n-1} - V_n &= L_0 \frac{dH_n}{dt}, \\
V_{n-1} - V_n &= L_1 \frac{dF_n}{dt}, \\
I_{n}^{(L)} - I_{n+1}^{(R)} &= C(V_n) \frac{dV_n}{dt} + \frac{V_n}{R},
\end{align}
where $F$ and $H$ represent the electric currents flowing through the passive and active inductances, respectively, as shown in Fig.~\ref{fig:numerical_scheme_unit_cell}.
Note that while the passive inductance is a conventional inductor, the active counterpart functions as an effective inductor, emulating the behavior of active operational amplifiers that introduce an inductance offset. We then use the rescaled time $t\rightarrow \omega_0 t$ with $\omega_0 = 1/\sqrt{C_0L_0}$, as well as the equations: 
\begin{equation}
    \begin{split}
        I_{n+1}^{(R)} &= H_{n+1} - F_{n+1} \\
        I_n^{(L)} &= H_n + F_n 
    \end{split},
\end{equation}
and arrive at the following system of equations:
\begin{equation}
    \begin{split}
        \frac{dH_n}{dt} &= a_1 \left(V_{n-1}-V_n\right) \\
        \frac{dF_n}{dt} &= \frac{b}{2} a_1 \left(V_{n-1}-V_n\right) \\
        \frac{dV_n}{dt}  &= \frac{1}{a_1} \frac{1}{\widetilde{C}(V_n)}\left(H_n + F_n - \left(H_{n+1}-F_{n+1}\right) - \frac{V_n}{R} \right) 
    \end{split}, 
\end{equation}
where
\begin{equation}
    a_1 = \frac{C_0}{L_0}, \quad \frac{b}{2} = \frac{L_0}{L_1}, \quad 2\gamma = \frac{1}{R} \sqrt{\frac{L_0}{R_0}}.
\end{equation}
We note that, by  applying a time derivative $d/dt$ to the third equation of the system above, we recover the second order differential equations of the rescaled variable with $y_n \rightarrow V_n$ discussed in the text.
The integration of the equations is done using a 4th-order Runge-Kutta method 
~\cite{HNW1993} and a time step $\Delta t=0.005$ to ensure high accuracy in our numerical simulations.
To simulate the experiments we drive the left end of the chain (i.e., $n=0$) using the boundary condition 
\begin{equation}
V_0= V_s(t) = A \sech^2 \left[\left(t - t_0\right)/\tau \right],
\end{equation}
where $\tau$ and $A$ are the width and amplitude of the pulse, respectively, and $t_0$ is the time where the pulse amplitude becomes maximum. 
At the opposite boundary, $n = N+1$, we apply open boundary conditions by enforcing gradient matching, such that $V_{N+1} = V_N$.
As a result, we have $H_{N+1} = F_{N+1} = 0$, representing a trivial current background due to $I_{N+1}^{(R)} = 0$.
A similar process is used for exciting the lattice from the opposite side.

\section{Electric non-reciprocal metamaterial with tunable on-site nonlinear resistance}

A modified discrete set of equations ins used to include nonlinear losses i.e.
\begin{equation}
    \frac{d^2y_n}{dt^2}  = a_{+} y_{n+1} - 2 y_n + a_{-}y_{n-1} - 2 \frac{d\left(\tilde{\gamma} y_n\right)}{dt} + \beta \frac{d^2y^2}{dt^2},
    \label{eq:ODEs_to_nonlinear_electric}
\end{equation}
with $\tilde{\gamma} = \gamma_0 - \gamma_1 y + \mathcal{O}\left(y^2\right)$ mimicking a voltage dependent resistance.
Such resistance relation can for instance be found in varistors electric elements.
Following the method above, we find the equation of motion in the continuum approximation in the weak nonlinear limit reads 
\begin{align}
        \frac{\partial^2 y}{\partial^2} - \frac{\partial^2 y}{\partial x^2} -  \frac{\partial^4 y}{\partial x^4} - \beta \frac{\partial \left( y^2\right)}{\partial t} \approx 
        -b \frac{\partial y}{\partial x} - \frac{b}{6}\frac{\partial^3 y}{\partial x^3} - 2\gamma_0\frac{\partial y}{\partial t} + 2\gamma_1 \frac{\partial \left( y^2\right)}{\partial t}.
\end{align}

In this equation we have also kept the third spatial derivative term and it is thus valid for higher values of $b$ than Eq.~\eqref{eq:continuum_small_b_01}.
We try to find solutions by assuming that both the left and right hand side are equal to zero.
We obtain for the left hand side,
\begin{equation}
    y(x,t) = \frac{3 (v^2 - 1)}{2\beta v^2} \sech^2 \left[\sqrt{3 (v^2 -1)} \left(x-vt\right)\right],
\end{equation}
and for the right hand counterpart, 

\begin{equation}
    y(x,t) = \frac{6\gamma_0v - 3b}{4\gamma_1v}\sech^2\left[\sqrt{\frac{6\gamma_0v-3b}{2b}}(x - vt)\right].
\end{equation}

It is readily seen that both the latter equation are satisfied if the nonlinear losses amplitude is such that 
$\gamma_1 \approx b\beta v$,
and
\begin{equation}
    \gamma_0 v + \frac{b}{2} \approx bv^2, \quad v \approx \frac{\gamma_0 + \sqrt{2b^2+\gamma_0^2}}{2b}
\end{equation}
with $b>2\gamma_0$ when requiring $v>1$.

\end{document}